# Exploring Electron Beam Induced Atomic Assembly via Reinforcement Learning in a Molecular Dynamics Environment

Rama K. Vasudevan,[1,a] Ayana Ghosh,[1,2] Maxim Ziatdinov,[1,2] and Sergei V. Kalinin[1,b]

[1] Center for Nanophase Materials Sciences and [2] Computational Sciences and Engineering Division, Oak Ridge National Laboratory, Oak Ridge, TN 37831

Atom-by-atom assembly of functional materials and devices is perceived as one of the ultimate targets of nanoscience and nanotechnology. While traditionally implemented via scanning probe microscopy techniques, recently it has been shown that the beam of a scanning transmission electron microscope can be used for targeted manipulation of individual atoms. However, the process is highly dynamic in nature and proceeds via a large number of weakly-understood individual steps. Hence, harnessing an electron beam towards atomic assembly requires automated methods to control the parameters and positioning of the beam in such a way as to fabricate atomic-scale structures reliably. Here, we create a molecular dynamics environment wherein individual atom velocities can be modified, effectively simulating a beam-induced interaction, and apply reinforcement learning to model construction of specific atomic units consisting of Si dopant atoms on a graphene lattice. We find that it is possible to engineer the reward function of the agent in such a way as to encourage formation of local clusters of dopants, whilst at the same time minimizing the amplitude of momentum changes. Inspection of the learned policies indicates that of fundamental importance is the component of velocity perpendicular to the material plane, and further, that the high stochasticity of the environment leads to conservative policies. This study shows the potential for reinforcement learning agents trained in simulated environments for potential use as atomic scale fabricators, and further, that the dynamics learned by agents encode specific elements of important physics that can be learned.

---

[a] vasudevanrk@ornl.gov
[b] sergei2vk@ornl.gov

Assembly of matter atom by atom towards the devices and materials with desired functionality is perceived as the ultimate goal of nanoscience and nanotechnology, as envisioned by some of the leading scientists of the 20th century.[1,2] Beyond the emergent and serendipitous applications in quantum computing, nanorobotics, and medicine, the ability to fabricate atomic scale structures is instrumental for physical research. As stated by Feynman, "what I cannot make I cannot understand." Correspondingly, capability to fabricate atomically defined structures and probe their functionalities is key to unlocking the fundamental physics of atomic world.

While perceived as largely speculative for most of the last century, the experiments by Don Eigler via Scanning Tunneling Microscopy demonstrated the potential of this method to create selected atomic structures[3,4] and heralded the explosive growth of nanoscience in the last thirty years. In accordance with the Feynman statement, many examples of fundamental studies such as quantum corrals,[5] molecular cascades,[6] quantum dots[7] and more have been demonstrated. Over last decade, combination of the surface chemistry and STM manipulation was established as enabling component of quantum computing infrastructure allowing for creation of phosphorus qubits.[8-10] Remarkably, while not having same degree of precision, many other atoms including oxygen and Sulphur can be manipulated.[11,12]

In the last decade, it was shown that sub-atomically focused electron beam in Scanning Transmission Electron Microscopy (STEM) can induced controllable changes in atomic structure on the atomic level. While electron beam damage was recognized as an issue since the earliest days of electron microscopy, the advances brought by aberration correction[13-15] illustrated that the effects of the energy transfer from the electron beam to the solid can be much more subtle, including activation of the set of chemical transformations and phase transitions,[16,17] lateral and rotational motion of atoms and molecular groups, and vacancy formation. It was immediately realized that control of these transformations can be used as a basis for atomic fabrication,[18] and direct controllable motion of Si atoms in graphene,[19,20] single-site doping,[19] and even formation of the homoatomic[21] and heteroatomic[22] molecular structures has been demonstrated.

However, the electron beam manipulation in STEM, as compared to probe-based manipulation in STM and non-contact Atomic Force Microscopy (nc-AFM), has a number of significant differences. First, STEM offers the potential for much higher throughput, since the intrinsic latencies in electron beam motion and image acquisition are much lower than that of mechanical probe in STM and noncontact-AFM. However, while the probe-induced chemistry in STM can be well-controlled, electron beams can induce a broad variety of beam-induced reactions, with associated difficulty of control. Both considerations necessitate the automatization of the process, i.e., incorporation of automated workflows controlling the microscope, to discover the cause-and-effect relationships between probe position and induced dynamic changes in the atomic structure, and further harness these towards controllable modification of solids.

Here, we explore the use of reinforcement learning (RL) for atomic manipulation implemented using a model molecular dynamics-run (MD) environment. The effect of the electron beam is modeled using deterministic energy momentum transfer to the selected atoms in the MD system. We show that utilizing RL, it is capable to train agents to perform atomic-level assembly in a very highly stochastic environment, and furthermore, inspect learned policies. The associated policy shifts are subtle, indicating the highly dynamic and somewhat unpredictable nature of the environment and explore how reward shaping affects the final policy. Some notes on extensibility of the process are discussed.

**Methods**

As a model system, we have utilized Si dopants in graphene with a Lennard-Jones potential given within the python-based atomic simulation environment (ASE).[23] Full details of the environment and model are given in the in Supplementary S1. Briefly, the MD simulations utilize a 64 atoms hexagonal graphene supercell where one or a few C atoms can be replaced by an Si dopant. Momenta can be imparted on the atoms in all three (x,y,z) dimensions. This sets the environment for the reinforcement learning formulation of the problem.

**Formulation of the RL Problem**

Prior to implementation of reinforcement learning on top of the molecular dynamics environment, we briefly discuss the fundamental RL concepts, namely the environment, state, action, and reward. The environment describes the system within which agent operates. In this case, this is molecular dynamics model. The state is defined as what is available or visible to the agent, and in this case, is the static projection of the surface, emulating the traditional high-angle annular dark field (HAADF) images. Here the state also includes the previous HAADF image, so that knowledge of the dynamic changes is inherently encoded in the shifts in atomic coordinates. The actions are the actions that agent can take, and here comprises the momentum transfer from the beam to the atomic unit emulating the beam positioning in the specified registry towards the selected nucleus and assuming deterministic momentum transfer. Finally, reward is an externally defined measure that can be tuned based on the desired structure and/or properties. Here, the ultimate target of the learning is the assembly of the desired atomic structure. Exact formulation of the reward function comprises an important part of RL algorithm that strongly affects the learning speed and will be discussed below.

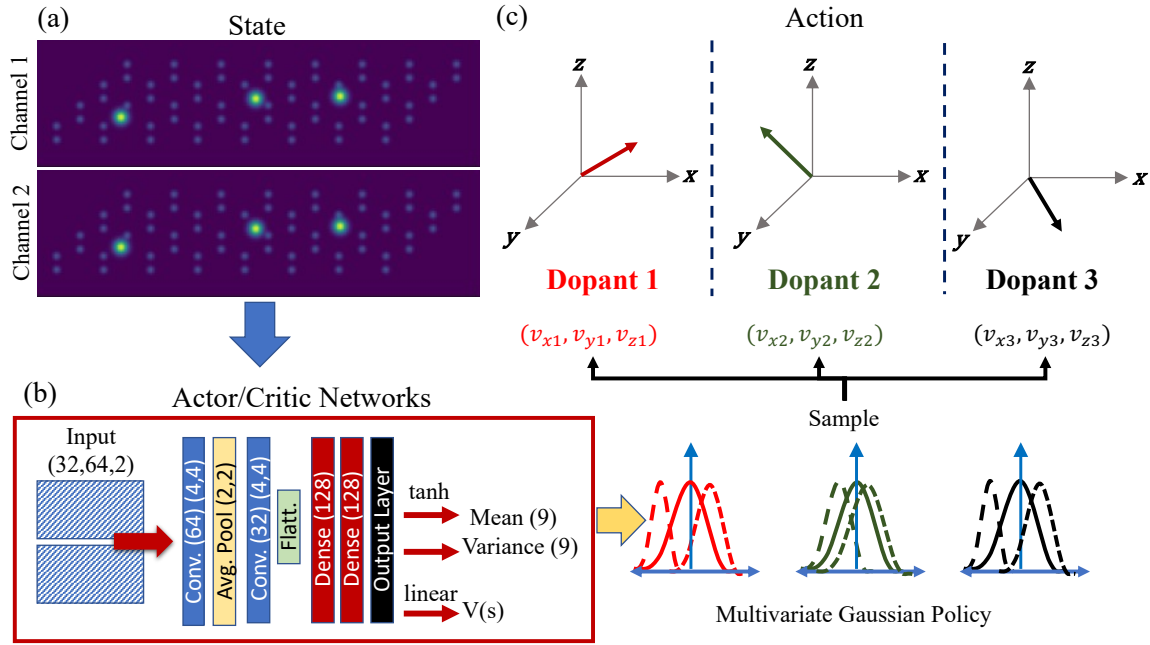

**Figure 1**: An example of the state utilized in the RL framework is shown in (a). The actor/critic network architecture is presented in (b). The multivariate gaussian as sampled based on the outputs of the networks and the velocity components of the three dopants are shown in (c).

An example of the state is shown in Fig. 1(a). The three Si dopant atoms are always initiated to be in the same position at the beginning of each episode. After 300 time steps, the HAADF image is returned to the agent. The added velocities or actions lead to change in position of atoms even after performing the task for a small time as compared to situations when no external energy is being introduced to the system for constant energy simulations. The agent's policy is a multivariate Gaussian policy, the mean and variance of which are parameterized by a neural network (Fig. 1(b)). This neural network, termed the 'actor' neural network, takes as input the HAADF images at time $t=t$, $t=t-1$, and passes the data through two convolutional layers and two fully connected layers, before outputting the mean and standard deviations for the multivariate gaussian policy. This same network structure is duplicated for the 'critic' network, which outputs the value of the state $V(s)$ (discussed below). Based on the outputs of the actor network, the multivariate gaussian is sampled, and then the momentum transfer to the three dopants is achieved via modification of the individual dopant's $x, y$ and $z$ components of velocity in the MD simulation, as shown in Fig. 1(c). The process then repeats until 6000 MD time steps are completed, signifying the end of the episode.

**Reward Function**

We aim to train policies that will assemble the three Si dopants as closely as possible, as has been shown experimentally. We define the reward function as

$$R = w_0 * d_S + w_1 * d_c + w_2 * A_{sum} \tag{1}$$

Where $w_0$, $w_1$ and $w_2$ are weight coefficients, $d_S$ defines the sum of (Euclidean) distances between the three Si dopants, $d_c$ is the sum of (Euclidean) distance of dopants to the center of the lattice, and $A_{sum}$ is the magnitude of the actions imparted on the atoms. The first part of the reward function deals with the desirability to create trimer-like structures, and this sets $w_0$ to be <0 (to penalize dopants moving away from each other). Similarly, $w_1$ is <0 given that we wish to penalize dopants agglomerating away from the center of the image. Finally, $w_2$ can be set to 0, in which case there is no penalty for large momentum transfers; we explore this parameter in detail below but note that it serves as a kind of regularization to limit large momentum transfers which can have undesirable effects (such as inducing more lattice instability, causing vacancies, etc.). Default values for $w_0 = -100$ and $w_1 = -300$. The reward is returned after every action, i.e., a total of 20 times per episode. Clearly, given the negative values of the weights, the maximum available reward for the agent is less than 0.

The goal of RL[24] is to learn a policy to maximize the cumulative (usually 'discounted) reward, i.e., to maximize the objective function $J$:

$$J(\pi) = \mathbb{E}\left[\sum_{t=0}^{\infty} \gamma^t r(s_t, a_t)\right] \quad (2)$$

where $\gamma$ is a discount factor between 0 and 1, actions $a_t$ are drawn from the policy $\pi(a_t|s_t)$ and the states $s_{t+1}$ are drawn from the environment, as a result of taking action $a_t$ at state $s_t$. Maximizing this expectation can be achieved via perturbing the policy parameters $\theta$ in the direction of the gradient of $J$ with respect to the policy parameters, and is given by the policy gradient formulation:

$$\nabla_\theta J(\pi) \approx \sum_{t=0}^{\infty} \nabla \log \pi(a_t|s_t; \theta) R_t \quad (3)$$

Where $R_t$ is the cumulative discounted return from time $t$, i.e., $R_t = \sum_{i=0}^{\infty} \gamma^i r(s_{t+i}, a_{t+i})$. Note that this formulation ensures that actions that affect future rewards are encouraged. Unfortunately, this formulation has very high variance in practice, and this variance can be substantially reduced via use of an appropriate baseline. One choice is the value function baseline, giving rise to the actor-critic algorithm:

$$\nabla_\theta J(\pi) \approx \sum_{t=0}^{\infty} \nabla \log \pi(a_t|s_t; \theta)(R_t - V^\pi(s_t)) \quad (3)$$

Where the 'Value' function $V^\pi(s_t) = \mathbb{E}\left[\sum_{i=0}^{\infty} \gamma^i r(s_{t+i}, a_{t+i})\right]$. This formulation is sometimes termed as 'advantage actor-critic' (A2C)[25], and intuitively expresses the idea that actions that provide greater advantage than the current expectation of the value of being in that state should be encouraged, and those that provide less returns than expected should be discouraged. This value function is estimated by a neural network termed the critic. A recent extension of the A2C approach is the so-called Stein variational policy gradient (SVPG) approach.[26, 27] More details of implementation can be found in Supplementary S2.

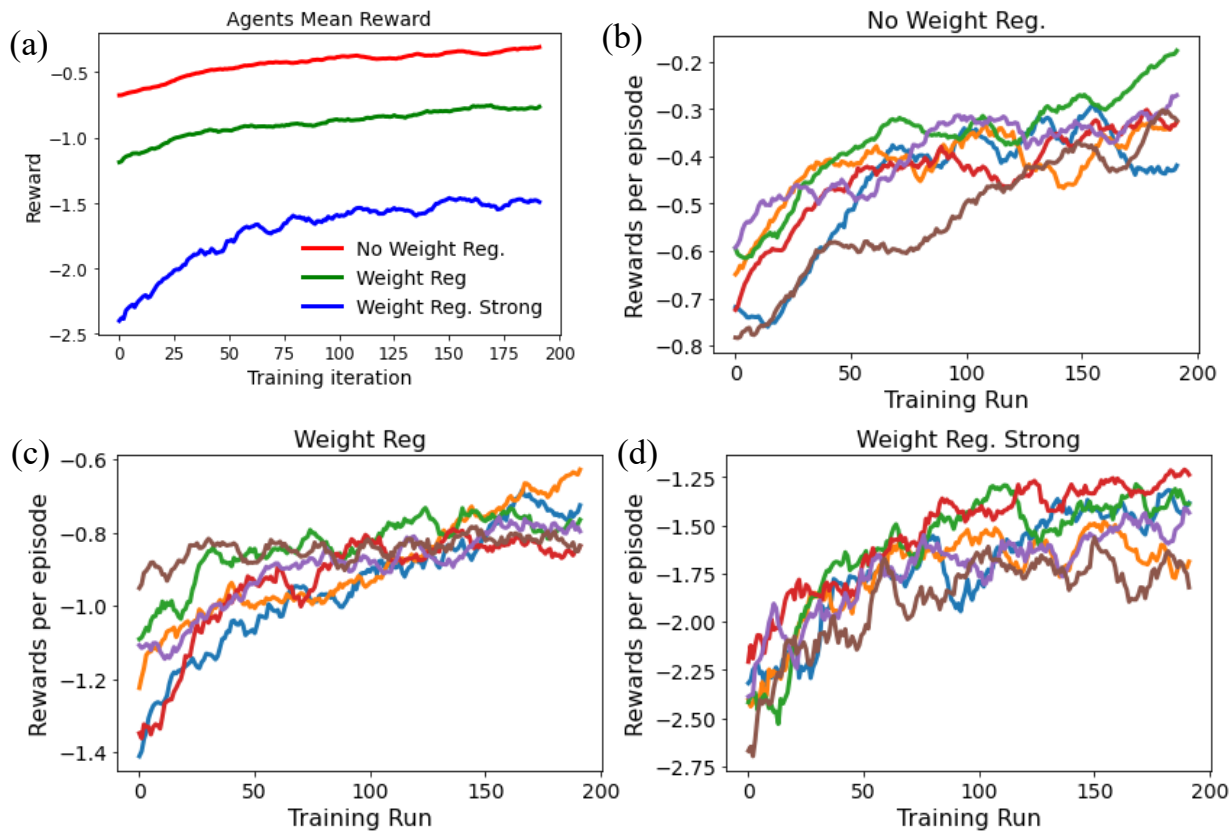

**Figure 2**. The mean of the training agents versus training iterations for three different weight regularization parameters are plotted in (a). The rewards as a function of training runs are shown in (b-d) for the corresponding regularization parameters. Different colors in (b-d) correspond to different agents. Smoothing of the training curves has been applied to each case with a window size of 10.

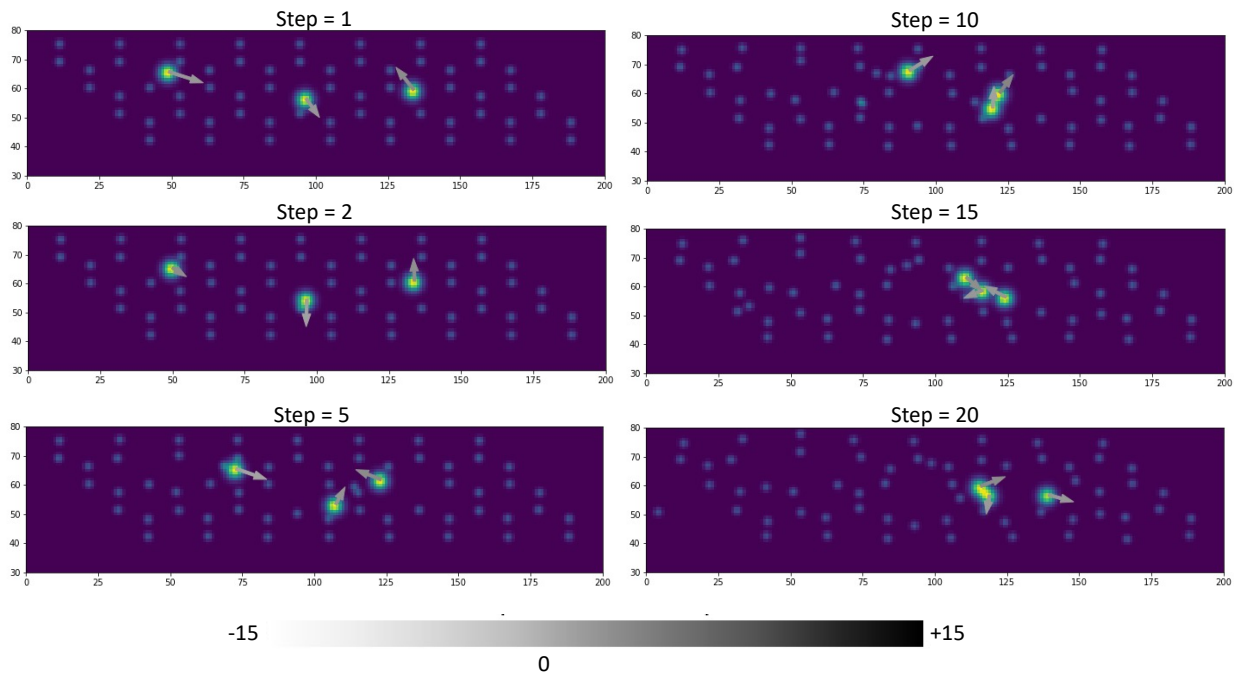

**Figure 3**. Example run using no weight regularization for different steps is shown here.

Results from the mean of the trained agents is shown in Fig. 2(a) for three different values of weight regularization parameter, $w_2$. Of note, all of the means show steady improvement through training, although it is evident that individual agents show non monotonic increases which is somewhat typical of training in RL. The reward values are not directly comparable due to the different reward functions for each of (b-d), but progress appears saturated at the high weight regularization, and near saturated for the other two scenarios.

Subsequently, we visualized the results of the highest performing agent in the case of $w_2 = 0$, i.e., no weight regularization. An example run is shown in Figure 3, which indicates the progression of states through episode step 1 to 20. Arrows indicate the actions taken at each step to each dopant atom, and the color of the arrow indicates the degree of '*z*' component to the velocity added. The agent shows strong success at moving the dopants together towards the center of the lattice, and this progress is highly repeatable across runs. This confirms that RL agents can be trained in this environment to adjust dopant momenta such that they agglomerate towards pre-chosen sites on the graphene lattice.

**Effect of Weight regularization**

To observe the effects of weight regularization, we trialed the trained agents with three different values of weight regularization parameter $w_2$ for five episodes and accumulated statistics of the actions performed. The violin plot of the actions for the three different agents are shown in Fig. 4(a), and clearly indicates that higher levels of weight regularization led to smaller changes to velocity, i.e., that this achieves what the reward function was designed to accomplish. It should be noted, however, that this did not in turn lead to the goal of reducing lattice disordering during the dopant-moving assembly process; the reason may be that rather than the overall magnitude of the velocity vector, it is the individual components of velocity that act to disrupt the lattice structure. This behavior can also be tied to an equidistribution of magnitudes of velocities that is considered at the initialization stage of the MD simulations, if the components are not mentioned individually, reflecting in the computed forces of individual atoms. We have also seen that specifying the velocity components individually as an action helps to retain the lattice structure. It is of course possible to alter the reward function such that we directly penalize the lattice disruption, however this can be more difficult in terms of design of reasonable metrics of lattice degradation and is not explored in this study. Moreover, note that in experiment, it has already been shown possible to move Si dopants in graphene without significant damage to the host lattice.[28] Finally, very strong weight regularization did result in decreased ability to agglomerate the dopants towards the center of the lattice, but this effect was less pronounced when the weight regularization was weaker.

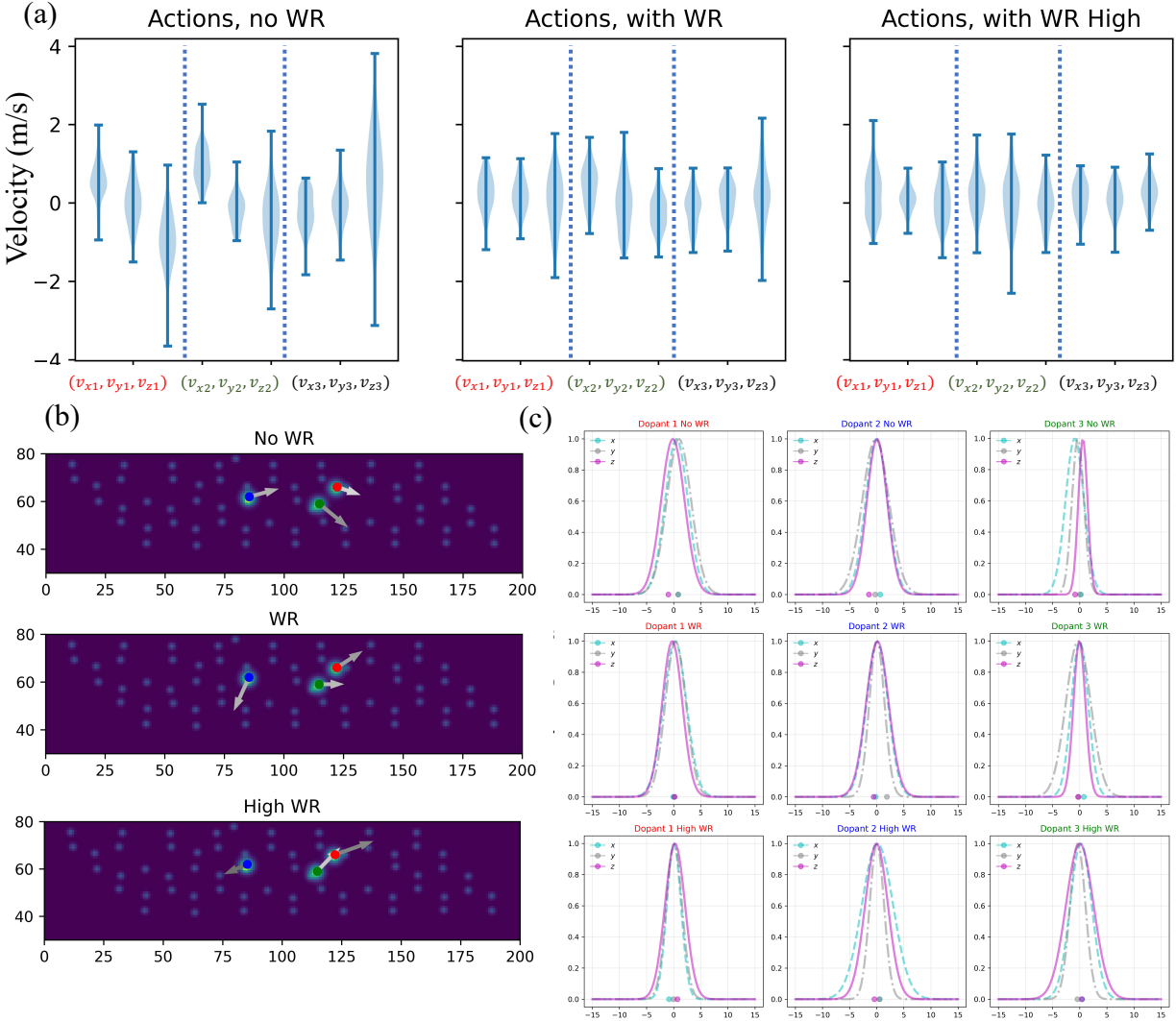

**Figure 4**: The actions for three different agents are shown as violin plots in (a), which were trained with different levels of weight regularization. The actions performed on a particular state and respective policies from actions are plotted in (b) and (c), respectively.

To obtain more insight into how the shape of the reward function affects the learned policies, we plot the policy for a randomly chosen state from the environment. The actions performed on this state by the three agents are shown in Fig. 4(b), and the respective policy from which the actions were sampled are plotted as Gaussians in Fig. 4(c). All three agents applied velocity vectors that were to the right, for the green and red dopant atoms, but differed in their preference to move the blue Si dopant. Inspection of the actual policies in Fig. 4(c) reveal several expected trends. The first is that the policies do not deviate much from ideal Gaussians, i.e., changes to velocity should be rather subtle. This may be due to the high stochasticity of the environment. Next, as the weight regularization increases, the variance of the policy reduces, in accordance with reducing the overall strength of the actions to increase the cumulative reward. Finally, and perhaps most importantly, the z-component of the velocity appears to be important because there is considerable weight on either side of the peak; this suggests that some small z-component of velocity is beneficial for the dopant motion. Similar conclusions were arrived to at in mechanistic analysis by Susi[20, 29-31] previously and is therefore very notable that the RL agent has, through only training, also discovered this mechanism.

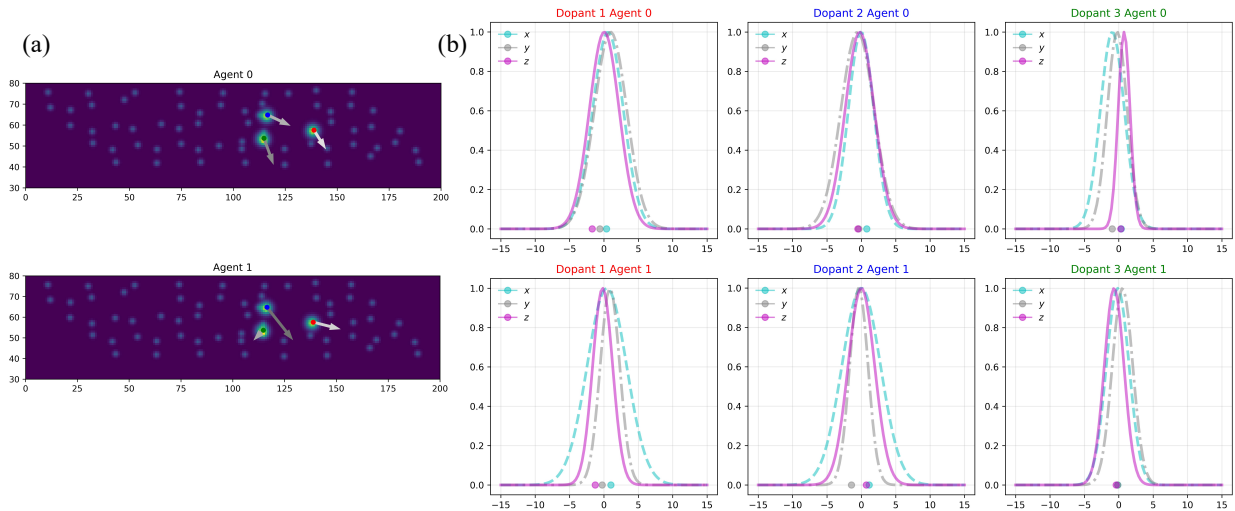

**Figure 5**: A random state (a) and the resultant policies generated by top two achieving agents are shown in (b).

**Ensemble training**

Finally, we note that the Stein variational policy gradient method enables training multiple agents in an ensemble. One of the advantages of such an approach is that this variance can improve the robustness of policies, i.e., learning multiple different methods to achieve the same target objective. An example of how the policies differ is shown in Fig. 5(a, b), where a random state is selected, and the two top-achieving agents are tested. Substantial differences in the policies are evident and are particularly evident for example for the momentum transfer to dopant 3 (green). This is further confirmed by inspecting the trained agents throughout the run (Supplementary S3) where small differences between the two agents are readily apparent.

**Summary**

To summarize, here we combine the reinforcement learning approach with molecular dynamics simulations to explore atomic structure fabrication via an electron beam. For the chosen model of beam interactions as momentum transfer, we found it is possible to engineer the momentum transfer in such a way as to deterministically move dopants towards desired locations and create agglomerations, in view of future potential of atomic-scale matter assembly via an automated approach. We explore the use of regularization in the reward function to temper the momentum transfer, and inspect multiple policies trained in ensemble to produce multiple methods of achieving the same target goal, potentially improving robustness.

This analysis illustrates the need for development of models of atomistic effects during beam solid interactions, including fast momentum transfer using classical knock-on mechanism and more subtle effects associated with local excitation of bonding electrons[32, 33] and hot carriers.[34] Notably, the dearth of theoretical models in this domain suggests that future exploration should target the co-navigation approaches, when the model is being refined along the experiment and in turn used to guide RL strategy. Given the extremely short latencies of the STEM experiment, this in turn necessitates development of edge computing capabilities bringing the MD processing to the edge of STEM experiment. Overall, this approach is expected both to enable atom-by-atom fabrication and in the process get insight into beam-induced chemical effects.

**Acknowledgements:**
This effort was performed and partially supported (S.V.K., R.K.V., M.Z.) at the Oak Ridge National Laboratory's Center for Nanophase Materials Sciences (CNMS), a U.S. Department of Energy, Office of Science User Facility and by U.S. Department of Energy, Office of Science, Office of Basic Energy Sciences Data, Artificial Intelligence and Machine Learning at DOE Scientific User Facilities program under Award Number 34532 (A.G.).

**Supplementary**
Supplementary information on the modeling environment, details of the implementation of the RL algorithm as well as videos of different runs of the trained agents are provided, and a link to an online Jupyter notebook (in Google Colab) is provided to reproduce the figures.

*Supplementary Information for*

Exploring Electron Beam Induced Atomic Assembly via Reinforcement Learning

in a Molecular Dynamics Environment

Rama K. Vasudevan,[1] Ayana Ghosh,[1,2] Maxim Ziatdinov,[1,2] and Sergei V. Kalinin[1]

[1]Center for Nanophase Materials Sciences and [2] Computational Sciences and Engineering Division, Oak Ridge National Laboratory, Oak Ridge, TN 37831

**Supplementary S1 – MD Environment**

As a model system, we utilized Si dopants in graphene with a Lennard-Jones potential given within the python-based atomic simulation environment (ASE).[1] For these MD simulations, a 64 atoms hexagonal graphene supercell, with lattice parameters (a=19.6 Å, b=9.8Å, c=20Å) is constructed and one of the C atoms is randomly replaced by a Si dopant atom. Periodic boundary conditions in all directions are imposed during each simulation. A Quasi-Netwon algorithm, namely BFGSLineSearch as implemented in ASE, is used to optimize the structure with 1000 iterations. These types of algorithms decide the positions of atoms at every iteration based on the forces and second derivative of total energy of the atoms. Once the geometry is optimized, a Velocity Verlet algorithm is employed to perform constant-energy MD simulations for 10,000 timesteps, 0.0032 fs of each timestep length.

In general, systems with lighter atoms and carbon bonds require a much smaller time step to prevent the system to become unstable. In general, systems with lighter atoms and carbon bonds require a much smaller time step to prevent the system to become unstable. In the femtoseconds range, even after performing the dynamics for a longer duration, the system does not break apart as compared to using picoseconds range. The changes in trajectory and the average properties of the dopant as well as other atoms can be traced once the simulation finishes. Moreover, here we are interested in testing the effect of added momentum to one of the atoms versus giving this external push to all the atoms in the system. For a longer timestep (>1 fs), this effect becomes challenging to track down once the simulation is completed since quantifying this effect necessitates the system evolution to be treated in a more controlled manner for all time steps. While a timestep as small as considered for these simulations and performed for 10,000 steps may not have significant changes in the absence of any external momentum added to the system, the scenario differs once the dopant is subjected to an external action and tries to equilibrate.

We explored the MD simulation environment by imparting different levels of momenta in the x and z directions and measuring the corresponding displacement of the Si dopant atom in the graphene lattice. Results are shown in Figure S1. Notably, the displacement to the dopant as a result of these momentum transfers is highly nonlinear as would be expected. Large transfers also lead to distortions of the surrounding lattice.

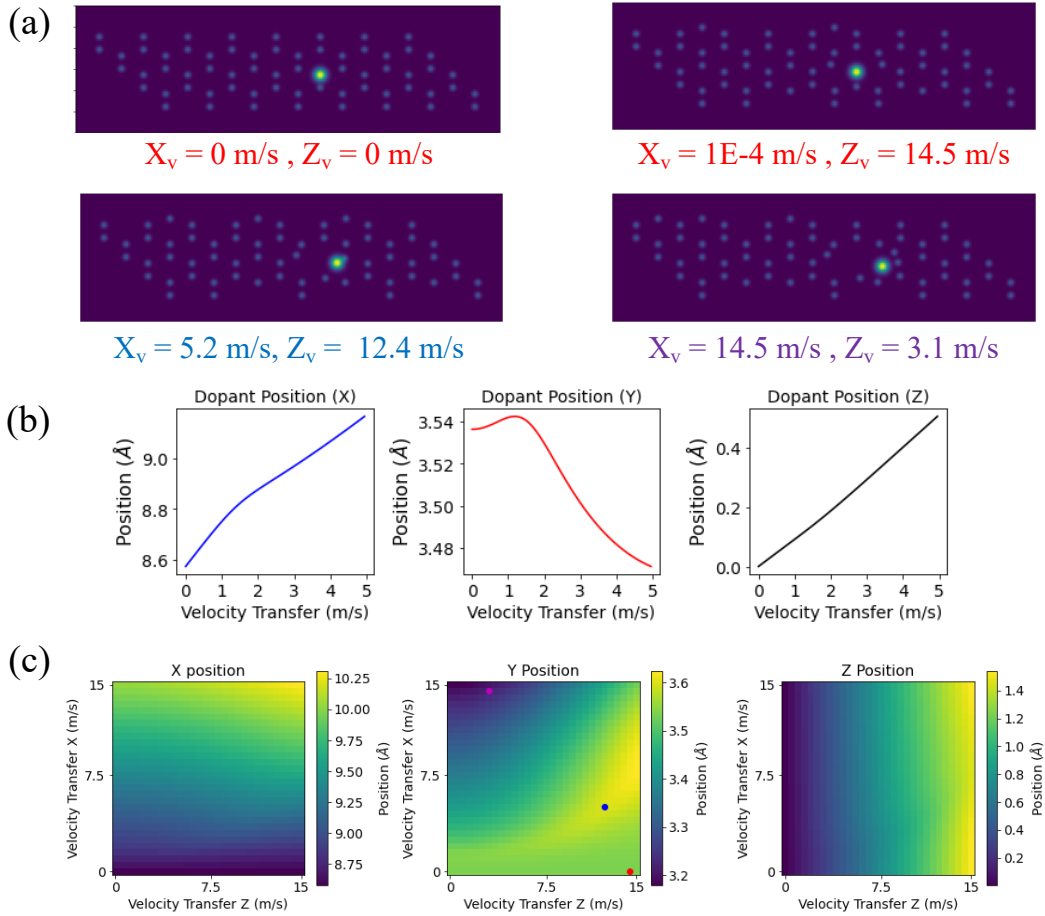

**Figure S1**: Graphene supercell used for MD simulations with one Si (highlighted) dopant for different *x* and *z* velocities are shown in **(a)**. The variations in *x, y, z* positions of the Si dopant atom as a function of velocity transfer and corresponding lattice distortions for a range of small to high initial momentum assigned to Si dopant atom are shown in **(b)** and **(c)**, respectively. End states of the parameters marked by the magenta, blue and red circles in the Y-Position panel in (c) are plotted in (a).

**Supplementary S2 – Implementation of Reinforcement Learning**

To train agents, the Stein variational policy gradient approach was utilized.[26, 27] Similar to equation (3), this is also a policy gradient approach, but instead of updating a single policy, the goal is to learn a distribution over policies that can maximize the objective function. This will likely lead to increased robustness and greater exploration, as more agents can operate in parallel and potentially learn different aspects, which can be shared with weight updates. We aim to train an ensemble of *n* agents each represented by a parameter vector, $\theta_i$ and the update step is given by

$$\Delta\theta_i \leftarrow \frac{1}{n}\sum_{j=1}^{n}[\nabla_{\theta_j}(\frac{1}{\alpha}J(\theta_j) + \log q_0(\theta_j))\,k(\theta_j,\theta_i) + \nabla_{\theta_j}k(\theta_j,\theta_i)] \quad (4)$$

where $\alpha$ is a hyperparameter that controls the degree of exploration (= 2.0), we set $\log q_0(\theta)$ is a flat prior and is set to 1, and the kernel function *k* used is the radial basis function and measures the 'distance' between the policies. It is immediately clear that when $\alpha$ is large, the gradient is less affected by individual gradients from the agents, and more emphasis is on the second term which encourages separation of the policies.

Conversely, setting alpha to be too small will discourage exploration. Here, agents were run in parallel to collect results from a single episode, before equation (4) was used to update the individual policies. Note that as with the original SVPG paper, the critic weights are not shared amongst agents and are updated independently via the traditional temporal difference (TD) error. Learning rates for the actor was 2E-4 and critic was 4E-4. Agents were trained on an Nvidia DGX-2 machine utilizing 16 GPUs and 40 core processor for a total of 200 episodes. Each episode consists of 6000 MD steps, and thus each run, where 6 agents were trained in parallel, consists of 36,000 MD steps per episode (or 7.2M MD steps in total).